\documentstyle[prl,aps,floats,epsf]{revtex}

\begin{document}
\draft

\twocolumn[\hsize\textwidth\columnwidth\hsize\csname
@twocolumnfalse\endcsname

\title{
Formation and rapid evolution of  
domain structure at phase transitions 
in slightly inhomogeneous  ferroelectrics and ferroelastics
}
\author{A.M. Bratkovsky$^{1}$ and A.P. Levanyuk$^{1,2}$}

\address{$^{1}$ Hewlett-Packard Laboratories, 1501 Page Mill Road, Palo
Alto, California 94304\\
$^{2}$ Departamento de F\'{i}sica de la Materia Condensada, C-III,
Universidad Aut\'{o}noma de Madrid, 28049 Madrid, Spain
}
\date{\today }
\maketitle

\begin{abstract}

We present the analytical study of stability loss and
evolution of domain structure in inhomogeneous
ferroelectric (ferroelastic) samples for exactly solvable models.
The model assumes a short-circuited ferroelectric capacitor 
(free ferroelastic)
 with two regions with slightly
different critical temperatures $T_{c1}>T_{c2}$,
where $T_{c1}-T_{c2} \ll T_{c1}, T_{c2}$. 
We show that even a tiny inhomogeneity like $10^{-5}$K
results in splitting the system into domains below the phase
transition temperature.
At $T<T_{c2}$ the domain width $a$  is proportional to 
$(T_{c1}-T)/(T_{c1} - T_{c2})$ and quickly increases with
lowering temperature. 
The minute inhomogeneities in $T_c$ may result from structural 
(growth) inhomogeneities, which are always present in real samples,
and a similar role can be played by inevitable temperature gradients.


\end{abstract}
\vskip 2pc ]

\section{Introduction}

The idea that the phase transition in electroded short-circuited
ferroelectric proceeds into homogeneous monodomain state\cite{Ginzburg49} is
very well known. Similar result also applies to free ferroelastic crystals.
However, it has {\em never} been observed. Surprisingly, both electroded
ferroelectrics and free ferroelastics do split into domains, although they
should not. The present paper aims to answer why.

It is generally assumed that in the finite non-electroded
ferroelectric samples the domain 
structure appears in order to reduce the depolarizing electric field if
there is a nonzero normal component of the polarization at the surface of
the ferroelectrics \cite{Ginzburg49,Kanzig52} (in complete analogy with
ferromagnets \cite{Landau35}), if the field cannot be reduced by either
conduction (usually negligible in ferroelectrics at low temperatures) or
charge accumulation from environment at the surface \cite{Jona62}. On the
other hand, in inhomogeneous ferroelastics (e.g. films on a substrate, or
inclusions of a new phase in a matrix) the elastic domain structure
accompanies the phase transition in order to minimize the strain energy, as
is well understood in case of martensitic phase transformations \cite
{Khachaturyan} and epitaxial thin films \cite{Roytburd76,BLprl2,BLstloss}.

In search for reasons of domain appearance in otherwise perfect electroded
samples, which is not yet understood, we shall discuss (i) a second order
ferroelectric phase transition in slightly inhomogeneous electroded sample
and (ii)\ a second order ferroelastic phase transition in slightly
inhomogeneous free sample. These problems have not been studied before. We
consider an exactly solvable case of a system, which has two slightly
different phase transition temperatures in its two parts. While the phase
transition occurs in the ``soft'' part of the system, the ``hard'' part may
effectively play a role of a ``dead'' layer \cite{BLprl1} and trigger a
formation of the domain structure in the ``soft'' part with fringe electric
fields (stray stresses) penetrating the ``hard'' part. One has to check this
possibility, but the behavior of the corresponding domain structure is
expected to be unusual: it should strongly depend on temperature since
further cooling transforms the ``hard'' part into a ``soft'' one, while the
first ``soft'' part becomes ``harder''. Since the inhomogeneity is small,
one might expect that the domains would quickly grow with lowering
temperature. We indeed find a rapid growth of the domain width linearly with
temperature in the case of slightly inhomogeneous short-circuited
ferroelectric and free ferroelastic. This behavior is generic and does not
depend on particular model assumptions. Generally, the inhomogeneous
ferroelectric systems pose various fundamental problems and currently
attract a lot of attention. In particular, {\em graded} ferroelectric films
and ferroelectric {\em superlattices} have been shown to have giant
pyroelectric \cite{graded} and unusual dielectric response\cite{FEsuper}.

\section{Phase transitions in slightly inhomogeneous ferroelectric }

We shall first consider the case of slightly inhomogeneous uniaxial
ferroelectric in short-circuited capacitor that consists of two layers with
slightly different critical temperatures, so that, for instance, a top part
``softens'' somewhat earlier than the bottom part does. We assume the easy
axis $z$ perpendicular to electrode plates, and make use of the Landau free
energy functional for given potentials on electrodes $\varphi _{a}$ (zero in
the present case)\cite{LLvol8} $\tilde{F}=F_{LGD}\left[ \vec{P}\right] +\int
dV\frac{E^{2}}{8\pi }-\sum_{a}e_{a}\varphi _{a},$ with 
\begin{eqnarray}
F_{LGD}[\vec{P}] &=&\sum_{p=1,2}\int dV\Bigl[\frac{A_{p}}{2}P_{z}^{2}+\frac{B%
}{4}P_{z}^{4}  \nonumber \\
&&+\frac{D}{2}\left( \nabla _{\perp }P_{z}\right) ^{2}+\frac{g}{2}\left(
\partial _{z}P_{z}\right) ^{2}+\frac{A_{\perp }}{2}\vec{P}_{\perp
}^{2}\Bigr],  \label{eq:lgd}
\end{eqnarray}
where $P_{z}$ $\left( \vec{P}_{\perp }\right) $is the polarization component
along (perpendicular to) the ``soft'' direction, index $p=1(2)$ marks the
top (bottom) part of the film: 
\begin{eqnarray*}
A_{1} &=&A,\qquad 0<z<l_{1}, \\
A_{2} &=&A+\delta A,\qquad -l_{2}<z<0.
\end{eqnarray*}
Here $A_{1(2)}=\alpha (T-T_{c1(2)})$ and $\delta A>0$ (meaning $%
T_{c2}<T_{c1})$. The constant $\alpha =1/T_{0},$ where $T_{0}\sim T_{at}$ $%
(T_{c})$ for displacive (order-disorder)\ type ferroelectrics, $T_{at}\sim
10^{4}-10^{5}$K is the characteristic atomic temperature.

The equation of state is $\delta F_{LGD}[\vec{P}]/\delta \vec{P}=\vec{E}%
=-\nabla \varphi ,$ where $\varphi $ is the electrostatic potential, or in
both parts of the film $p=1,2:$%
\begin{eqnarray}
E_{z} &=&-\partial _{z}\varphi =A_{p}P_{z}+BP_{z}^{3}-D\nabla _{\perp
}^{2}P_{z}-g\partial _{z}^{2}P_{z},  \label{eq:Ez} \\
\vec{E}_{\perp } &=&A_{\perp }\vec{P}_{\perp },  \label{eq:Eperp}
\end{eqnarray}
These equations should be solved together with the Maxwell equation, ${\rm %
div}(\vec{E}+4\pi \vec{P})=0,$ or 
\begin{equation}
\left( \partial _{z}^{2}+\epsilon _{a}\nabla _{\perp }^{2}\right) \varphi
=4\pi \partial _{z}P_{z},  \label{eq:div1}
\end{equation}
where the dielectric constant in the plane of the film is $\epsilon
_{a}=1+4\pi /A_{\perp }.$

\subsection{Loss of stability$.$}

We shall now find conditions for loss of stability of the paraelectric phase
close to $T_{c1}$ with respect to inhomogeneous polarization. The stability
loss corresponds to appearance of a non-trivial solution to linearized
equations of equilibrium. Indeed, at the brink of instability the system is
in neutral equilibrium, defined by linear terms. We are looking for a
nontrivial solution in a form of the ''polarization wave'', 
\begin{equation}
P_{z},\varphi \propto e^{ikx}.
\end{equation}
We shall check later that the stability will be lost for the wave vector $%
kl_{1}\gg 1$ while the scale of change of $P_{z}$ with $z$ is $l_{1}$ so
that $\nabla _{\perp }^{2}P_{z}=k^{2}P_{z}\gg g\partial _{z}^{2}P_{z}\sim
P_{z}/l_{1}^{2},$ and the last term in the right-hand side of (\ref{eq:Ez})\
should be dropped. Going over to Fourier harmonics indicated by the
subscript $k$, we obtain for the Poisson equation: 
\begin{equation}
\varphi _{k}^{\prime \prime }-\epsilon _{a}k^{2}\varphi _{k}=4\pi
P_{zk}^{\prime },  \label{eq:fiP}
\end{equation}
where the prime indicates derivative $($ $f^{\prime }\equiv df/dz,$ $%
f^{\prime \prime }\equiv d^{2}f/dz^{2})$. We can exclude $P_{zk}$ with the
use of the linearized equation of state (\ref{eq:Ez}), which gives 
\begin{equation}
-\varphi _{k}^{\prime }=(A_{p}+Dk^{2})P_{zk}.  \label{eq:lineq}
\end{equation}
Substituting this into (\ref{eq:fiP}), we obtain $\varphi _{k}^{\prime
\prime }-\frac{\epsilon _{a}k^{2}\left( A_{p}+Dk^{2}\right) }{4\pi }\varphi
_{k}=0,$ where we have used $\left| A+Dk^{2}\right| /4\pi \ll 1,$ which is
always valid in ferroelectrics. We shall see momentarily that the nontrivial
solution appears only when $A_{1}+Dk^{2}<0,$ while $A_{2}+Dk^{2}>0.$ The
resulting system is 
\begin{eqnarray}
\varphi _{1k}^{\prime \prime }+\chi _{1}^{2}k^{2}\varphi _{1k} &=&0,
\label{eq:f1} \\
\varphi _{2k}^{\prime \prime }-\chi _{2}^{2}k^{2}\varphi _{2k} &=&0,
\label{eq:f2}
\end{eqnarray}
where $\chi _{1}^{2}=-\epsilon _{a}\left( A_{1}+Dk^{2}\right)/4\pi ,
$ $\chi _{2}^{2}=\epsilon _{a}\left( A_{2}+Dk^{2}\right)/4\pi.$
The corresponding solutions satisfying the boundary conditions for
electroded surfaces ($\varphi =0$ at $z=l_{1},$ $l_{2})$ read 
\begin{eqnarray}
\varphi _{1k} &=&F\sin \chi _{1}k(z+l_{1}), \\
\varphi _{2k} &=&G\sinh \chi _{2}k(z-l_{2}).
\end{eqnarray}
The boundary condition at the interface ($z=0)$ reads as 
\begin{equation}
\frac{\varphi _{1k}^{\prime }}{A_{1}+Dk^{2}}=\frac{\varphi _{2k}^{\prime }}{%
A_{2}+Dk^{2}},  \label{eq:bcprime}
\end{equation}
where we have used $\left| A_{1}+Dk^{2}\right| /4\pi \ll 1.$ We obtain\ from
Eqs. (\ref{eq:f1})-(\ref{eq:bcprime}) the condition for a nontrivial
solution 
\begin{equation}
\chi _{1}\tan \chi _{1}kl_{1}=\chi _{2}\tanh \chi _{2}kl_{2},
\label{eq:disp}
\end{equation}
which has a homogeneous solution $k=0$ and the inhomogeneous solution with $%
k \neq 0$ (\ref{eq:kcel}), hence we have to determine which one is actually
realized. The inhomogeneous solution is easily found for $\chi
_{2}kl_{2}\gtrsim 1,$ where $\tanh $ can be replaced by unity. Close to the
transition $\chi _{2}/\chi _{1}\gg 1,$ and the solution is 
\begin{equation}
\chi _{1}kl_{1}=\frac{\pi }{2}\frac{\chi _{2}kl_{1}}{1+\chi _{2}kl_{1}}%
\approx \frac{\pi }{2},
\end{equation}
when $\chi _{2}kl_{1}\gg 1.$ This gives 
the condition of stability loss in the form
$|A|=Dk^{2}+\frac{\pi ^{3}}{\epsilon
_{a}k^{2}l_{1}^{2}}.$ There is no solution for $\chi _{1}^{2}<0.$ The minimal
value of $A$ for the nontrivial solution (the actual onset of
instability, if the transition with $k=0$ does not occur earlier) is defined by 
\begin{eqnarray}
k_{c} &=&\left( \frac{\pi ^{3}}{\epsilon _{a}Dl_{1}^{2}}\right)
^{1/4}\approx \frac{\pi ^{3/4}}{\epsilon _{a}^{1/4}}\frac{1}{\sqrt{%
d_{at}l_{1}}},  \label{eq:kcel} \\
|A|_{c} &=&2Dk_{c}^{2}=\frac{2\pi ^{3/2}D^{1/2}}{\epsilon _{a}^{1/2}l_{1}}%
\approx \frac{2\pi ^{3/2}}{\epsilon _{a}^{1/2}}\frac{d_{at}}{l_{1}},
\label{eq:Acel}
\end{eqnarray}
where we have introduced the ``atomic'' size $d_{at}\sim \sqrt{D}$
comparable to the lattice parameter. We obtain the corresponding tiny shift
in the critical temperature [see estimates below Eq.(\ref{eq:dTc})] $%
T_{c1}-T_{c}\sim T_{0}d_{at}/\epsilon _{a}^{1/2}l_{1}.$ Hence, the system
looses its stability with respect to an inhomogeneous structure
very quickly below the bulk transition temperature. It
is readily checked that the assumptions we used to obtain the solution are
easily satisfied. Indeed, $\chi _{2}kl_{2}\gtrsim 1$ and $\chi _{2}kl_{1}\gg
1$ both correspond to approximately the same condition when $l_{1}\sim
l_{2}:\delta A\gg \frac{4}{\pi ^{1/2}\epsilon _{a}^{1/2}}\frac{d_{at}}{l_{1}}%
,$ meaning that the difference between $T_{c}$ should be larger than the
shift of $T_{c}$.

Now we have to determine when the transition into inhomogeneous state occurs
prior to a loss of stability with respect to a {\em homogeneous}
polarization. The homogeneous loss of stability corresponds to $A=A_{h}$
found from 
\begin{equation}
A_{h}l_{1}+\left( A_{h}+\delta A\right) l_{2}=0.
\end{equation}
For the inhomogeneous state to appear first, there must be $A_{c}>A_{h},$ or 
$\delta A>\frac{\pi ^{3/2}(l_{1}+l_{2})}{\epsilon _{a}^{1/2}l_{1}}\frac{%
d_{at}}{l_{1}}.$ This means that very {\em tiny inhomogeneity} in the sample
is enough to split it into the domain structure, 
\begin{equation}
T_{c1}-T_{c2}=T_{0}\frac{\pi ^{3/2}(l_{1}+l_{2})}{\epsilon _{a}^{1/2}l_{1}}%
\frac{d_{at}}{l_{1}},  \label{eq:dTc}
\end{equation}
which, for a film $1%
\mathop{\rm mm}%
$ thick, is estimated as $T_{at}\frac{d_{at}}{\epsilon _{a}^{1/2}l_{1}}%
\lesssim \epsilon _{a}^{-1/2}(10^{4}-10^{5})10^{-7}$K$=(10^{-3}-10^{-2}){\rm %
K}$ for displacive systems, and $T_{c}\frac{d_{at}}{\epsilon _{a}^{1/2}l_{1}}%
\lesssim \left( 10^{-5}-10^{-4}\right) ${\rm K} for order-disorder systems.
Certainly, such a small temperature and/or compositional inhomogeneity
exists in all usual experiments.

\subsection{Domain structure at $T_{c2}<T<T_{c1}$ $(A<0,${\em \ }$A+\protect%
\delta A>0).$}

After stability loss the resulting ''polarization wave'' quickly develops
into a domain structure, as we shall now demonstrate. The notion of the
domain can be applied when the domain width $a=\pi /k_{c}$ becomes
comparable and larger than the domain wall thickness $W\sim \sqrt{D/|A|}.$
The relation $W\lesssim a$ gives [see Eqs.(\ref{eq:kcel},\ref{eq:Acel})] 
\begin{equation}
|A|\gtrsim \left( \frac{D}{\pi \epsilon _{a}}\right) ^{1/2}\frac{1}{l_{1}}%
\approx \frac{d_{at}}{\epsilon _{a}^{1/2}l_{1}}\ll 1.  \label{eq:dTdomains}
\end{equation}
This is the same tiny temperature interval where the present scenario
unfolds, and system quickly goes over into domain state well above the lower
transition temperature $T_{c2}$, if it is larger than the value
defined by Eq.~(\ref{eq:dTc}). 

In the region below $T_{c1}$ where the domain structure forms (as shown
above, it occupies most of the temperature interval $T_{c1}-T_{c2})$, we can
use the linearized equation of state 
\begin{equation}
E_{z}=(A+3BP_{01}^{2})(P_{z}-P_{01})=-2A(P_{z}-P_{01}),
\end{equation}
where $\left| P_{01}\right| =\sqrt{-A/B}$ is the spontaneous polarization in
the top layer, which gives $P_{z1}=P_{01}+\frac{1}{2|A|}E_{z}$, $P_{z2}=%
\frac{1}{A_{2}}E_{z},$ for the top and bottom layers, respectively. In this
case the equation for the potential $\varphi $ (\ref{eq:div1}) reduces to a
standard Laplace equation $\left( \epsilon _{c}\partial _{z}^{2}+\epsilon
_{a}\nabla _{\perp }^{2}\right) \varphi =0,$ with the boundary condition 
\begin{equation}
\epsilon _{c1}\partial _{z}\varphi _{1}-\epsilon _{c2}\partial _{z}\varphi
_{2}=4\pi P_{01}(x),
\end{equation}
where $\epsilon _{c1}=1+2\pi /|A|,$ $\epsilon _{c2}=1+4\pi /A_{2}.$

\begin{figure}[t]
\epsfxsize=2.8in \epsffile{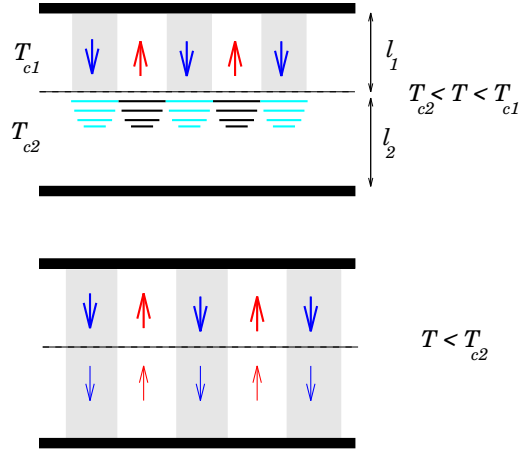}
\caption{ Schematic of the domain structure with the period $2a$ in
inhomogeneous ferroelectric film of the thickness $l_{1}+l_{2}$. Top and
bottom layers have slightly different critical temperatures $T_{c1}>T_{c2}$, 
$T_{c1}-T_{c2}\ll T_{c1},T_{c2}$. Slightly below $T_{c1}$ the top layer
splits into domains with electric fringe field propagating into the bottom
layer (fringe field shown as the hatched area in the top panel). The domains
persist and evolve below $T_{c2}$ when both layers exhibit a ferroelectric
(or ferroelastic) transition (bottom panel). }
\label{fig:fringe}
\end{figure}

The spontaneous polarization in the top layer alternates from domain to
domain as $P_{01}(x)=\pm \left| P_{01}\right| \equiv \pm \sqrt{-A/B}.$ We
are looking for a solution in a form of a domain structure with a period $%
T=2a$ (Fig. \ref{fig:fringe}), 
\begin{equation}
P_{01}(x)=\sum_{k}P_{01k}e^{ikx},\hspace{0.2in}\varphi (x)=\sum_{k}\varphi
_{k}e^{ikx},  \label{eq:fik}
\end{equation}
with $k\equiv k_{n}=2\pi n/T=\pi n/a,$ $n=\pm 1,\pm 2,...$ Going over to the
Fourier harmonics, we can write the Laplace equations for both parts of the
film as 
\begin{eqnarray}
\epsilon _{c1}\varphi _{1k}^{\prime \prime }-\epsilon _{a}k^{2}\varphi _{1k}
&=&0,  \label{eq:lap1} \\
\epsilon _{c2}\varphi _{2k}^{\prime \prime }-\epsilon _{a}k^{2}\varphi _{2k}
&=&0,  \label{eq:lap2}
\end{eqnarray}
with the boundary conditions at the interface $z=0$ 
\begin{equation}
\varphi _{1k}=\varphi _{2k},\qquad \epsilon _{c1}\varphi _{1k}^{\prime
}-\epsilon _{c2}\varphi _{2k}^{\prime }=4\pi P_{01k}  \label{eq:bc01}
\end{equation}
The corresponding {\em electrostatic} (stray) field part of the energy is
found as \cite{BLprl1} 
\begin{equation}
\tilde{F}_{es}=\frac{1}{2}\int d{\cal A}\sigma _{s}\varphi \left( z=0\right)
,
\end{equation}
where $\sigma _{s}$ is the density of bound charge at the interface ,
corresponding to {\em only} the spontaneous part of the polarization $%
P_{01}(x),$ and integration goes over the area ${\cal A}$\ between two parts
of the film. We calculate this expression by going over to Fourier expansion
(\ref{eq:fik}) and using the fact that in the present geometry $\sigma
_{s}(x)=-P_{01}(x)$ (and, therefore, its Fourier component $\sigma
_{sk}=-P_{01k}),$%
\begin{equation}
\frac{\tilde{F}_{es}}{{\cal A}}=\sum_{k>0}\frac{4\pi |P_{01k}|^{2}}{kD_{k}},
\label{eq:Fes01}
\end{equation}
\begin{equation}
D_{k}=\epsilon _{a}^{1/2}\left[ \epsilon _{c1}^{1/2}\coth \sqrt{\frac{%
\epsilon _{a}}{\epsilon _{c1}}}kl_{1}+\epsilon _{c2}^{1/2}\coth \sqrt{\frac{%
\epsilon _{a}}{\epsilon _{c2}}}kl_{2}\right] ,  \label{eq:Dk}
\end{equation}
with $k=\pi n/a,$ $n=1,2,...,$ similar to \cite{BLef}. Note that here $%
P_{01k}=2\left| P_{01}\right| /i\pi n,$ $n=2j+1,$ $j=0,1,...$ and zero
otherwise. Adding the surface energy of the domain walls, we obtain the free
energy of the domain pattern 
\begin{equation}
\frac{\tilde{F}}{{\cal A}}=\frac{\gamma _{1}l_{1}}{a}+\frac{16P_{01}^{2}a}{%
\pi ^{2}}\sum_{j=0}^{\infty }\frac{1}{\left( 2j+1\right) ^{3}D_{2j+1}},
\label{eq:Ft01}
\end{equation}
where $D_{n}=D_{k_{n}}.$ Not very close to $T_{c1}$ the argument of $\coth $
is $\sqrt{\frac{\epsilon _{a}}{\epsilon _{c1}}}kl_{1}\gtrsim 1$ even for the
smallest $k=\pi /a$ what is checked by the subsequent result (Eq.\ref{eq:a1})%
$,$ so that $D_{k}=\epsilon _{a}^{1/2}\left( \epsilon _{c1}^{1/2}+\epsilon
_{c2}^{1/2}\right) .$ Minimizing the free energy, we find the domain width 
\begin{equation}
a=\left[ \frac{\pi ^{2}\epsilon _{a}^{1/2}\left( \epsilon
_{c1}^{1/2}+\epsilon _{c2}^{1/2}\right) }{14\zeta (3)}\Delta _{1}l_{1}\right]
^{1/2},  \label{eq:a1}
\end{equation}
where $\Delta _{1}\equiv \gamma _{1}/P_{01}^{2}=d_{at}|A|^{1/2}$ is the
characteristic microscopic length, and $d_{at}\equiv \frac{2^{3/2}}{3}D^{1/2}
$ is comparable to\ a lattice spacing (``atomic'' length scale)$.$ The
expression (\ref{eq:a1}) is valid when $\sqrt{\frac{\epsilon _{a}}{\epsilon
_{c1}}}kl_{1}\gtrsim 1,$ or $|A|\gtrsim 2d_{at}/\left( \pi \epsilon
_{a}^{1/2}l_{1}\right) ,$ meaning that one has to be below $T_{c}$ by a tiny
amount $T_{c1}-T\gtrsim T_{0}d_{at}/\left( \epsilon _{a}^{1/2}l_{1}\right) ,$
estimated earlier. Note that close to $T_{c1}$ one obtains for the domain
width 
\begin{equation}
a=a_{K}\equiv \left[ \frac{\pi ^{5/2}\epsilon _{a}^{1/2}}{7\sqrt{2}\zeta (3)}%
d_{at}l_{1}\right] ^{1/2},  \label{eq:aKel}
\end{equation}
and this value does {\em not} depend on temperature. We shall formally refer
to this result as the Kittel domain width.

Incidentally, close to $T_{c2}$ the domain width is $a\approx \left[ \frac{%
\pi ^{2}\epsilon _{a}^{1/2}\epsilon _{c2}^{1/2}}{14\zeta (3)}\Delta _{1}l_{1}%
\right] ^{1/2}\propto \epsilon _{c2}^{1/4},$ which formally diverges $%
\propto \left( T-T_{c2}\right) ^{-1/4}.$ However, in the vicinity of $T_{c2}$
the induced polarization in the formerly ``hard'' part has about the same
value as the spontaneous polarization in the ``soft'' part, $P_{z2}\approx
P_{01}.$ Then the equation of state in the bottom part becomes strongly
non-linear, since the cubic term is much larger than the linear term, $%
BP_{z2}^{3}\approx BP_{01}^{3}=AP_{01}\approx AP_{z2}\gg A_{2}P_{z2}$, in
the equation of state (since $A\gg A_{2}$ close to $T_{c2})$, so the
response of the bottom layer does not actually soften in this region. In
this case our derivation does not apply, but it is practically certain that
the domain structure in the vicinity of $T_{c2}$ would evolve continuously
upon cooling, Fig. \ref{fig:aak}.

\subsection{Domain structure at low temperatures ($T<T_{c2},${\em \ }$A<0,$%
{\em \ }$A+\protect\delta A<0).$}

When the system is cooled to below the critical temperature $T_{c2}$, a
spontaneous polarization $\left| P_{02}\right| =\sqrt{-A_{2}/B}$ also
appears in the bottom layer. The domain structure simultaneously develops in
the whole crystal with domain walls running parallel to the ferroelectric
axis through the whole crystal (if they were discontinuous at the interface
between the two parts of the crystal this would have created a large
depolarizing electric field). The electrostatic energy requires a solution
of the same Laplace equations (\ref{eq:lap1}) and (\ref{eq:lap2}), only the
boundary condition (\ref{eq:bc01})\ would now read 
\begin{equation}
\epsilon _{c1}\varphi _{1k}^{\prime }-\epsilon _{c2}\varphi _{2k}^{\prime
}=4\pi (P_{01k}-P_{02k}),
\end{equation}
where $\epsilon _{c1(2)}=1+2\pi /|A_{1(2)}|\approx 2\pi /|A_{1(2)}|.$ Note
that the density of the bound charge at the interface, corresponding to this
discontinuity of spontaneous polarization, is now $\sigma
_{k}=-(P_{01k}-P_{02k}).$ Therefore, we immediately obtain for the total
free energy of the structure, analogously to the previous case (\ref{eq:Ft01}%
), 
\begin{eqnarray}
\frac{\tilde{F}}{{\cal A}} &=&\frac{P_{01}^{2}\Delta
_{1}l_{1}+P_{02}^{2}\Delta _{2}l_{2}}{a}  \nonumber \\
&&+\frac{16(P_{01}-P_{02})^{2}a}{\pi ^{2}}\sum_{j=0}^{\infty }\frac{1}{%
\left( 2j+1\right) ^{3}D_{2j+1}},  \label{eq:Ft12}
\end{eqnarray}
where $\Delta _{1(2)}=d_{at}\sqrt{|A_{1(2)}|}.$ Not very close to $T_{c2}$
we would have $\sqrt{\frac{\epsilon _{a}}{\epsilon _{c2}}}kl_{2}\gtrsim 1$
even for the smallest value of $k=\pi /a$ which enables us to replace $\coth 
$ by unity. The minimum of the free energy $\tilde{F}$ is achieved for the
domain width 
\begin{eqnarray}
a &=&\frac{1}{1-P_{02}/P_{01}}  \nonumber \\
&\times &\left[ \frac{\pi ^{2}\epsilon _{a}^{1/2}\left( \epsilon
_{c1}^{1/2}+\epsilon _{c2}^{1/2}\right) }{14\zeta (3)}\left( \Delta
_{1}l_{1}+\Delta _{2}l_{2}\frac{P_{02}^{2}}{P_{01}^{2}}\right) \right]
^{1/2}.
\end{eqnarray}
Close to the critical point $T_{c2}$ the domain width formally behaves as $%
a\propto \epsilon _{c2}^{1/4}\propto (T_{c2}-T)^{-1/4}$, as found just above 
$T_{c2}$ before. The same argument indicates though that our derivation does
not apply in this region, but non-linearity should not cause a substantial
change in the domain structure.

\begin{center}
\begin{figure}[t]
\epsfxsize=2.4in \epsffile{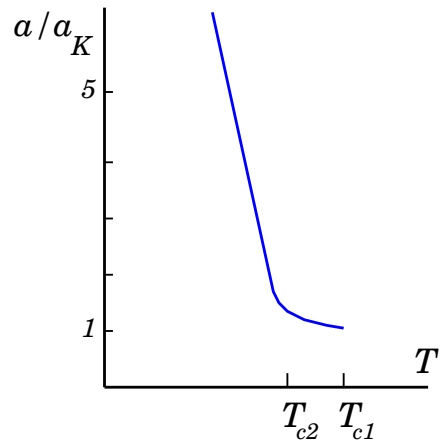}
\caption{ The domain width in slightly inhomogeneous ferroelectric or
ferroelastic in the units of $a_{K}$, the Kittel width (\ref{eq:aKel}). $%
a=a_{K}$ when the domain structure sets in at $T\approx T_{c1}$, and then it
grows linearly with the temperature to large values $a\gg a_{K}$. }
\label{fig:aak}
\end{figure}
\end{center}

With lowering the temperature to the region where $|A|\gg \delta A,$ we will
have $P_{02}/P_{01}=\sqrt{(A+\delta A)/A}\approx 1+\delta A/2A,$ so that $%
1-P_{02}/P_{01}\approx 2|A|/\delta A\gg 1$ becomes a large prefactor. Note
that in this region $\epsilon _{c1}\approx \epsilon _{c2}=2\pi /|A|,$ $%
\Delta _{1}\approx \Delta _{2}=d_{at}\sqrt{|A|},$ and the domain width
evolves as 
\begin{equation}
a=\frac{|A|}{\delta A}\left[ \frac{2^{5/2}\pi ^{5/2}\epsilon _{a}^{1/2}}{%
7\zeta (3)}d_{at}(l_{1}+l_{2})\right] ^{1/2},  \label{eq:domainwidth}
\end{equation}
It becomes much larger than the Kittel width, $\frac{a}{a_{K}}=2^{3/2}\left( 
\frac{l_{1}+l_{2}}{l_{1}}\right) ^{1/2}\frac{T_{c1}-T}{T_{c1}-T_{c2}}\gg 1,$
growing linearly with lowering temperature (Fig. \ref{fig:aak}). For large
periods of the domain structure Eq.(\ref{eq:domainwidth}) becomes
inapplicable because the $\coth $ in the formula for $D_{k}$ (\ref{eq:Dk})\
cannot be replaced by unity, and this corresponds to $|A|\simeq \left(
\delta A\right) ^{2}\frac{l_{2}}{d_{at}}$. If we assume that the difference
between the critical temperatures in the both parts of the system is, for
example, just $T_{c1}-T_{c2}=0.1$K. Since $|A|=(T_{c1}-T)/T_{0}$ and $\delta
A=(T_{c1}-T_{c2})/T_{0},$ we see that in 1mm thick film $(l/d_{at}\sim
10^{7})$ the expression for the domain structure period, Eq.(\ref
{eq:domainwidth}), is valid at least in the region 
\begin{equation}
T_{c1}-T\lesssim \frac{l_{2}}{d_{at}}\frac{\left( T_{c1}-T_{c2}\right) ^{2}}{%
T_{0}}.
\end{equation}
This interval is $1-10$K for displacive and $50-100$K for order-disorder
systems.

It follows from the qualitative analysis of the expression for the electric
energy, that the domain width $a$ will keep growing with lowering
temperature beyond this range to sizes much larger than the Kittel width (%
\ref{eq:aKel}), because the system quickly moves into the region $|A|\gg
\delta A.$ This result is rather natural, since in this limit the relative
difference between two parts of the system diminishes, and the system
approaches the limit of a uniform free sample, which transforms into a
monodomain state (i.e. $a=\infty ).$

\section{Inhomogeneous ferroelastic phase transitions}

Very scenario applies in a case of slightly inhomogeneous proper
ferroelastic in spite of some differences in the mathematics. Consider a
ferroelastic slab with slightly different phase transition temperatures, $%
T_{c1}>T_{c2},$ in its two parts of a comparable size. In such a situation,
the ''hard'' part will play a role of a rigid substrate for the top ``soft''
part of the sample at temperatures slightly below $T_{c1}$, and the sample
will split into domains. The emerging domain structure should strongly
evolve with temperature, since the bottom part of the film would also become
''soft'' at $T=T_{c2}$ slightly below $T_{c1}.$

\subsection{Loss of stability $\left( T\approx T_{c1}\right) $}

\begin{figure}[t]
\epsfxsize=3.2in \epsffile{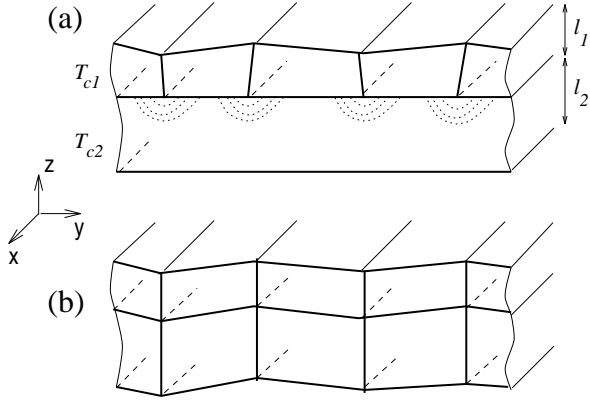 }
\caption{ Schematic of the domain structure in
inhomogeneous ferroelastic film of the thickness $l_{1}+l_{2}$ with
``soft'' in-plane strain $u_{xy}$. 
Top and bottom layers have slightly different critical temperatures
$T_{c1}>T_{c2}$,  
$T_{c1}-T_{c2}\ll T_{c1},T_{c2}$. (a) Slightly below $T_{c1}$ the top layer
splits into domains  with fringe elastic field near the interface
$z=0$ (schematically shown by the dotted lines).
(b) The domains persist and evolve below $T_{c2}$ when both layers exhibit a 
ferroelastic phase transition (bottom panel). }
\label{fig:domel}
\end{figure}

We assume that the film is perpendicular to the $z$-axis, occupies the space 
$-l_{2}<z<l_{1},$ and is characterized by the $u_{xy}$ (in-plane) component
of the strain tensor as the order parameter. The ``hard'' shear modulus
equals $\mu $ in both parts of the film, while the ``soft'' modulus
corresponds to the $u_{xy}$ component of strain. We shall consider a
situation when the system consists of two layers with slightly different
critical temperatures, Fig.~\ref{fig:domel}.
Thus, the Landau thermodynamic potential has the form 
\begin{eqnarray}
F &=&\sum_{p=1,2}\int dV[2A_{p}u_{xy}^{2}+2D\left( \nabla u_{xy}\right)
^{2}+Bu_{xy}^{4}  \nonumber \\
&&+\mu \left( u_{ik}^{2}-2u_{xy}^{2}\right) ]  \label{eq1}
\end{eqnarray}
where $A_{1}\equiv A=\alpha \left( T-T_{c1}\right) $, $A_{2}=\alpha \left(
T-T_{c2}\right) \equiv A+\delta A,$ with $\alpha ,D,\mu >0$ positive
constants, and $\delta A>0$ corresponding to $T_{c2}<T_{c1}.$ Thus, the top
layer of the system ''softens'' at $T_{c1}$ $\left( A=0\right) $ while the
other part of the system remains ''hard''. Note that $A$ designates now an
elastic modulus and not reciprocal dielectric susceptibility as in the
previously described case of ferroelectric. We designate the two parameters
by the same letter to underline the similarities in the corresponding
formulas.

The equation of state in each part $p$\ of the system is $\sigma _{ik}=\frac{%
1}{2}\delta F/\delta u_{ik},$ $i\neq k:$ 
\begin{eqnarray}
\sigma _{xy}^{p} &=&2(A_{p}-D\nabla ^{2})u_{xy}+2Bu_{xy}^{3},  \label{eq2} \\
\sigma _{xz}^{p} &=&2\mu u_{xz},\quad p=1,2,  \label{eq:sxz}
\end{eqnarray}
where $p=1$ ($2)$ corresponds to the part $0<z<l_{1}$ ($-l_{2}<z<0).$

The treatment of the stability loss is analogous to that of Ref. \cite
{BLprl2} and we omit some preliminary discussion presented in that earlier
paper. To find the inhomogeneous part of the {\em displacement} vector $%
u_{x} $ (or, equivalently, $u_{y}$ ) at the phase transition one should
satisfy the equations of local equilibrium, $\partial \sigma _{ik}/\partial
x_{k}=0,$ which in the present case read 
\begin{equation}
\frac{\partial \sigma _{xy}}{\partial y}+\frac{\partial \sigma _{xz}}{%
\partial z}=0.  \label{eq3}
\end{equation}
We shall use the Fourier expansion for the displacement vector 
\begin{equation}
u(y,z)=\int u_{k}\left( z\right) \exp \left( iky\right) dk  \label{eq4}
\end{equation}
and find the first appearance of the non-trivial solution for $u$ for a
given $k$ wavevector. We then determine the $k$ where the instability sets
in first, and this will be the point of the stability loss of the symmetric
phase.

We obtain the following equations for the displacement with the use of Eqs.(%
\ref{eq3}),(\ref{eq2}) 
\begin{eqnarray}
\frac{d^{2}u_{k}}{dz^{2}}-\frac{A_{1k}}{\mu }k^{2}u_{k} &=&0,\quad 0<z<l_{1};
\label{eq:film} \\
\frac{d^{2}u_{k}}{dz^{2}}-\frac{A_{2k}}{\mu }k^{2}u_{k} &=&0,\quad
-l_{2}<z<0,  \label{eq:sub}
\end{eqnarray}
where $A_{1k}=A+Dk^{2}$ and $A_{2k}=A+\delta A+Dk^{2}.$ At the free surfaces 
$\left( z=l_{1},-l_{2}\right) $ the boundary condition reads $\sigma
_{xz}=0, $ which is equivalent to $du_{k}(z)/dz=0$. In addition, the
displacement $u_{k}\left( z\right) $ and the stress $\sigma _{xz}\left(
z\right) $ should be continuous at the interface $z=0.$

Let us first consider the case of $A_{1k}<0$, $A_{2k}>0,$ which would
correspond, as we will see shortly, to a loss of stability of the paraphase.
The solution of Eqs. (\ref{eq:film}), (\ref{eq:sub}) is

\begin{eqnarray}
u_{k}(z) &=&F\cos \chi _{1}k(l_{1}-z),\quad 0<z<l_{1};  \label{eq8} \\
u_{k}(z) &=&G\cosh \chi _{2}k(z+l_{2}),\quad -l_{2}<z<0,
\end{eqnarray}
where $\chi _{1}^{2}=-A_{1k}/\mu =(-A-Dk^{2})/\mu $ and $\chi
_{2}^{2}=A_{2k}/\mu =(A+\delta A+Dk^{2})/\mu .$

The condition of existence of a non-trivial solution, which we obtain from
the boundary conditions, looks exactly the same as in the above case of
ferroelectrics, Eq.(\ref{eq:disp}).The subsequent analysis is also similar,
and we obtain a condition for an existence of a nontrivial solution 
\begin{equation}
|A|=\frac{\pi ^{2}\mu }{4k^{2}l_{1}^{2}}+Dk^{2},
\end{equation}
so the minimal value of $|A|=|A|_{c},$ when the solution first appears,
corresponds to 
\begin{eqnarray}
k_{c} &=&\sqrt{\frac{\pi }{2l_{1}d_{at}}},  \label{eq:kc} \\
|A|_{c} &=&2Dk_{c}^{2}=\frac{\pi \mu d_{at}}{l_{1}},  \label{eq:Ac}
\end{eqnarray}
where we have introduced the characteristic ``atomic'' length scale $%
d_{at}\sim \left( D/\mu \right) ^{1/2},$ which is comparable to unit cell
size. The corresponding shift of the
critical temperature is very small. The coefficient $\alpha $ in (\ref{eq1}%
)\ is $\mu /T_{0},$ where $T_{0}\sim T_{at}$ in the case a displacive, and $%
\sim T_{c}$ in the case of order-disorder phase transition. Then, from Eq.(%
\ref{eq:Ac}), 
\begin{equation}
T_{c}=T_{c1}-\frac{\pi \mu d_{at}}{\alpha l_{1}}\approx T_{c1}-T_{0}\frac{%
\pi d_{at}}{l_{1}},  \label{eq:dTc1}
\end{equation}
which is practically the same estimate, as for ferroelectrics, with the same
(by the order of magnitude) values for the displacive and order-disorder
phase transitions and the same condition for transition into inhomogeneous
instead of a homogeneous state.

\subsection{Domain structure in the top layer at $T_{c2}<T<T_{c1}$ ($A<0,$ $%
A+\protect\delta A>0$)}

We consider next the domain structure below $T_{c1}$ in a state with the
spontaneous strain $u_{xy}^{0}$. One can apply the notion of the domain
structure when the domain wall thickness is much smaller than the domain
width. This condition is fulfilled just below $T_{c1}$ by a tiny amount
given by the same small parameter $d_{at}/l_{1}$ as in the case of the
ferroelectrics, Eq.(\ref{eq:dTdomains}). As in the previous case, one can
apply the notion of the domain structure practically in the whole interval $%
T_{c1}-T_{c2}$, if the transition indeed proceeds into inhomogeneous state.
Within this interval one can use the linearized equation of state for the
top layer, obtained by expanding the free energy (\ref{eq1}) about the
spontaneous deformation, 
\begin{eqnarray}
\sigma _{xy}^{1} &=&2M_{1}(u_{xy}-u_{xy}^{0}),\qquad 0<z<l_{1}  \label{eq15}
\\
u_{xy}^{0} &\equiv &u_{0}=\pm \left( -A/B\right) ^{1/2},  \label{eq:uspont}
\\
\sigma _{xy}^{2} &=&2M_{2}u_{xy},\qquad -l_{2}<z<0,  \label{eq:s2tc2plus}
\end{eqnarray}
where $M_{1}\equiv -2A>0$ is $\ll \mu $ when the system is close to the
transition (``soft'' modulus), $M_{2}=A+\delta A>0.$ In both parts $\sigma
_{xz}$ is given by Eq.(\ref{eq:sxz}).

We shall assume that all the domains have the same width\cite{BLprl2}, which
we will find by minimizing the sum of the elastic energy and the (surface)\
energy of the domain walls. We consider a stripe-like domain structure in
the top layer with the spontaneous strain $u_{xy}^{0}(y,z)=\pm u_{0}$ with
the period $2a.$ There would be no stresses in the {\em free} top layer if $%
u_{0}^{2}=-A/B.$ We have to find the displacements $u_{x}(y,z)\equiv u\left(
y,z\right) $ appearing after the top layer experienced a phase transition.
The equation of equilibrium (\ref{eq3}) takes the form 
\begin{eqnarray}
M_{1}\frac{\partial ^{2}u}{\partial y^{2}}+\mu \frac{\partial ^{2}u}{%
\partial z^{2}} &=&2M_{1}\frac{\partial u_{xy}^{0}}{\partial y},\quad
0<z<l_{1}  \label{eq17} \\
M_{2}\frac{\partial ^{2}u}{\partial y^{2}}+\mu \frac{\partial ^{2}u}{%
\partial z^{2}} &=&0,\quad -l_{2}<z<0.
\end{eqnarray}
Since the domain pattern is periodic, the elastic displacements may be
represented as a Fourier series 
\begin{equation}
u\left( y,z\right) =\sum_{k}u_{k}\left( z\right) \exp \left( iky\right) ,%
\text{\qquad }k\equiv k_{n}=\frac{\pi n}{a},  \label{eq:fourie}
\end{equation}
where $n=\pm 1,\pm 2,\ldots $ After solving the resulting system of ordinary
differential equations with the above conditions one finds 
\begin{equation}
u_{k}(z)=u_{0}R_{k}\Bigl[\frac{\eta _{2}\cosh \eta _{1}k(z-l_{1})}{\eta
_{1}\sinh \eta _{1}kl_{1}\coth \eta _{2}kl_{2}+\eta _{2}\cosh \eta _{1}kl_{1}%
}-1\Bigr]  \label{eq:u1}
\end{equation}
at $0<z<l_{1},$ and 
\begin{equation}
u_{k}(z)=-\frac{u_{0}R_{k}\eta _{1}\cosh \eta _{2}k(z+l_{2})}{\eta _{1}\cosh
\eta _{2}kl_{2}+\eta _{2}\coth \eta _{1}kl_{1}\sinh \eta _{2}kl_{2}}
\label{eq:u2}
\end{equation}
at $-l_{2}<z<0.$ where $R_{k}=4/k^{2}a,$ $k=\pi \left( 2r+1\right) /a,$ $%
r=0,\pm 1,...$ and $\eta _{i}=\sqrt{M_{i}/\mu },$ $i=1,2,$ The elastic
energy is found with the use of the formula \cite{Mura} 
\begin{equation}
F_{el}=-\frac{1}{2}\int \sigma _{ij}u_{ij}^{0}dV=-\int \sigma
_{xy}u_{xy}^{0}dV  \label{eq19}
\end{equation}
with the result for the elastic (stray) energy per unit area of the film$:$%
\begin{equation}
\frac{F_{stray}}{\mu u_{0}^{2}{\cal A}}=\frac{16\eta _{1}\eta _{2}a}{\pi ^{3}%
}\sum_{j=0}^{\infty }\frac{1}{\left( 2j+1\right) ^{3}}\frac{1}{D_{j}}
\label{eq:fstc1}
\end{equation}
with $D_{j}=\eta _{1}\coth \eta _{2}k_{j}l_{2}+\eta _{2}\coth \eta
_{1}k_{j}l_{1}.$ To find the equilibrium domain width, we have to add the
energy of the domain walls 
\begin{equation}
F_{dw}/A=\gamma _{1}l_{1}/a=\mu u_{0}^{2}\Delta _{1}l_{1}/a,  \label{eq:Fdw1}
\end{equation}
where \cite{BLprl2} 
\begin{eqnarray}
\gamma _{1} &=&\frac{8\sqrt{2}D^{1/2}\left| A\right| ^{3/2}}{3B}\equiv \mu
u_{0}^{2}\Delta _{1}, \\
\Delta _{1} &\equiv &\frac{8\sqrt{2}D^{1/2}\left| A\right| ^{1/2}}{3\mu }%
\equiv d_{at}\sqrt{\left| A\right| /\mu }  \label{eq:Del1}
\end{eqnarray}
and we have introduced the microscopic length scale $\Delta _{1},$ with $%
d_{at}\equiv \frac{8\sqrt{2}}{3}\left( D/\mu \right) ^{1/2}$ is, once more,
of the order of the unit cell size. Note that the actual domain walls exist
only in the top layer, which underwent a ferroelastic transition, with the
stray displacement field penetrating into the bottom ``rigid'' part of the
sample.

The equilibrium domain width is found from the total free energy 
\begin{equation}
\frac{F_{tot}}{\mu u_{0}^{2}{\cal A}}=\frac{F_{stray}}{\mu u_{0}^{2}{\cal A}}%
+\frac{\Delta _{1}l_{1}}{a}  \label{eq:ftc1}
\end{equation}
with the stray energy from (\ref{eq:fstc1}). Assuming $\pi \eta
_{1(2)}l_{1(2)}/a\gtrsim 1$ (to be checked later), we replace all $\coth $
in (\ref{eq:fstc1})\ by unity and easily obtain for the domain width 
\begin{equation}
a=\sqrt{\frac{\pi ^{3}\Delta _{1}l_{1}}{14\zeta \left( 3\right) }\frac{\eta
_{1}+\eta _{2}}{\eta _{1}\eta _{2}}}.
\end{equation}
Slightly below $T_{c1}$ one has $\eta _{1}\ll \eta _{2}$%
\begin{equation}
a=a_{K}\equiv \sqrt{\frac{\pi ^{3}\Delta _{1}l_{1}}{14\zeta \left( 3\right) }%
\frac{1}{\eta _{1}}}=\sqrt{\frac{\pi ^{3}}{2^{3/2}7\zeta \left( 3\right) }%
d_{at}l_{1}},  \label{eq:aK}
\end{equation}
the limiting value which does {\em not} depend on temperature close to
transition. We shall formally call this a Kittel period for the elastic
domain structure, and the system, as we have shown, looses stability and
quickly sets in the domain structure with this period, which is independent
of the temperature close to the phase transition. As in
the case of ferroelectrics, the period of the domain structure formally
increases close to $T_{c2},$ but this conclusion is not reliable because the
nonlinear effects in the former ``hard'' layer should be taken into account
in this region.

\subsection{Ferroelastic domain structure at low temperatures ($T<T_{c2},$ $%
A<0,$ $A+\protect\delta A<0$)}

We consider next the domain structure not very close to the phase transition.
There the domain wall
width is much smaller than the width of the domains and one can use the
linearized equation of state in both top and bottom parts of the film,
obtained by expanding the free energy (\ref{eq1}) about the spontaneous
deformation, 
\begin{eqnarray}
\sigma _{xy} &=&2M_{1}(u_{xy}-u_{xy}^{0}),\qquad 0<z<l_{1}  \label{eq:s1tc2}
\\
\sigma _{xy} &=&2M_{2}(u_{xy}-w_{xy}^{0}),\qquad -l_{2}<z<0,
\label{eq:s2tc2}
\end{eqnarray}
where $M_{1}\equiv -2A$ is $\ll \mu $ (``soft'' modulus), $M_{2}=-2(A+\delta
A)$ [note the change in the $M_{2}$ value below $T_{c2}$]. Similarly to
previous case $u_{xy}^{0}\equiv u_{0}=\pm \left( -A/B\right) ^{1/2},$ $%
w_{xy}^{0}\equiv w_{0}=\pm \left[ -(A+\delta A)/B\right] ^{1/2}.$

We have to find the inhomogeneous displacements in the film $%
u_{x}(y,z)\equiv u\left( y,z\right) $. For the film the equation of
mechanical equilibrium (\ref{eq3}) takes the form 
\begin{eqnarray}
M_{1}\frac{\partial ^{2}u}{\partial y^{2}}+\mu \frac{\partial ^{2}u}{%
\partial z^{2}} &=&2M_{1}\frac{\partial u_{xy}^{0}}{\partial y},\qquad
0<z<l_{1}, \\
M_{2}\frac{\partial ^{2}u}{\partial y^{2}}+\mu \frac{\partial ^{2}u}{%
\partial z^{2}} &=&2M_{2}\frac{\partial w_{xy}^{0}}{\partial y},\qquad
-l_{2}<z<0.
\end{eqnarray}
We look for a solution in the same periodic form (\ref{eq:fourie}) as
earlier with the result 
\begin{equation}
u_{k}(z)=R_{k}\left[ \frac{\left( u_{0}-w_{0}\right) \eta _{2}\cosh \eta
_{1}k(z-l_{1})}{\eta _{1}\sinh \eta _{1}kl_{1}\coth \eta _{2}kl_{2}+\eta
_{2}\cosh \eta _{1}kl_{1}}-u_{0}\right] ,
\end{equation}
for $0<z<l_{1},$ and 
\begin{equation}
u_{k}(z)=-R_{k}\left[ \frac{\left( u_{0}-w_{0}\right) \eta _{1}\cosh \eta
_{2}k(z+l_{2})}{\eta _{1}\cosh \eta _{2}kl_{2}+\eta _{2}\coth \eta
_{1}kl_{1}\sinh \eta _{2}kl_{2}}-w_{0}\right]
\end{equation}
for $-l_{2}<z<0.$ The elastic (stray) energy per unit area of the film is
now found by integrating over both parts of the film, since now a
spontaneous strain exists in both of them: 
\begin{eqnarray}
\frac{F_{stray}}{{\cal A}} &=&\frac{16\eta _{1}\eta _{2}\mu
(u_{0}-w_{0})^{2}a}{\pi ^{3}}  \nonumber \\
&&\times \sum_{j=0}^{\infty }\frac{1}{\left( 2j+1\right) ^{3}}\frac{1}{\eta
_{1}\coth \eta _{2}k_{j}l_{2}+\eta _{2}\coth \eta _{1}k_{j}l_{1}},
\label{eq:Fs2}
\end{eqnarray}
where $\eta _{i}=\sqrt{M_{i}/\mu },$ $i=1,2,$ with $M_{1}=-2A,$ $%
M_{2}=-2(A+\delta A)$ and $k_{j}=\pi (2j+1)/a.$ Closer to $T=T_{c2}$ from
below this expression becomes similar to that for the previous case, since $%
w_{0}\rightarrow 0.$ To find the total free energy one has to add the energy
of the domain walls 
\begin{equation}
\frac{F_{dw}}{{\cal A}}=\frac{\mu u_{0}^{2}\Delta _{1}l_{1}+\mu
w_{0}^{2}\Delta _{2}l_{2}}{a},
\end{equation}
where $\Delta _{2}=d_{at}\left| \frac{A+\delta A}{\mu }\right| ^{1/2}$,
while $\Delta _{1}$ is given by (\ref{eq:Del1}). The equilibrium period of
the structure is 
\begin{equation}
a=\frac{1}{1-w_{0}/u_{0}}\sqrt{\frac{\pi ^{3}\left[ \Delta _{1}l_{1}+\Delta
_{2}l_{2}\left( w_{0}^{2}/u_{0}^{2}\right) \right] }{14\zeta \left( 3\right) 
}\frac{\eta _{1}+\eta _{2}}{\eta _{1}\eta _{2}}},  \label{eq:a2}
\end{equation}
in the same approximation as before, $\pi \eta _{2}l_{2}/a\gtrsim 1,$ which
enables us to replace $\coth $ by unity in (\ref{eq:Fs2}).

We find in the vicinity of $T_{c2},$ where $\eta _{2}\equiv \eta _{2}^{-}=%
\sqrt{-2(A+\delta A)/\mu }\ll \eta _{1},$ $\Delta _{2}\ll \Delta _{1},$ $%
w_{0}\ll u_{0}$), the equilibrium domain width 
\begin{equation}
a=\sqrt{\frac{\pi ^{3}\Delta _{1}l_{1}}{14\zeta \left( 3\right) }\frac{1}{%
\eta _{2}^{-}}}\propto \frac{1}{\sqrt{\eta _{2}^{-}}},  \label{eq:atc2m}
\end{equation}
We see that the period of the domain structure formally diverges when
one is approaching $T_{c2}$ from below but this behavior will be
modified by the nonlinear effects.

Let us check the behavior of the domain width at temperatures deep into the
ferroelastic region for both parts of the film, where $|A|\gg \delta A.$
There $\eta _{2}\approx \eta _{1},$ $\Delta _{2}\approx \Delta _{1}$ and $%
1-w_{0}/u_{0}\approx \delta A/2|A|\ll 1,$ and we obtain 
\begin{equation}
a=\frac{2|A|}{\delta A}\sqrt{\frac{\pi ^{3}}{7\sqrt{2}\zeta \left( 3\right) }%
d_{at}(l_{1}+l_{2})}.  \label{eq:avak}
\end{equation}
We see that far below the temperature where a spontaneous strain sets in the
whole system, the period of the domain structure grows with respect to the
Kittel period of initial domain structure $a_{K},$ Eq.~(\ref{eq:aK}), as 
\begin{eqnarray}
\frac{a}{a_{K}} &=&\frac{2^{3/2}|A|}{\delta A}\left( \frac{l_{1}+l_{2}}{l_{1}%
}\right) ^{1/2}  \nonumber \\
&=&2^{3/2}\left( \frac{l_{1}+l_{2}}{l_{1}}\right) ^{1/2}\frac{T_{c1}-T}{%
T_{c1}-T_{c2}}\gg 1.  \label{eq:biga}
\end{eqnarray}
Since the period is linearly growing with lowering temperature, $a\propto
|A|,$ and becomes very large, $a\gg a_{K}$, and the replacement of the $%
\coth $ by unity becomes unjustified. The condition of applicability of Eq.(%
\ref{eq:biga}) is the same as for Eq.(\ref{eq:domainwidth})\ in the case of
ferroelectrics.

\section{Summary}

Summarizing, in an electroded ferroelectric or free ferroelastic sample with
a tiny inhomogeneity of either the critical temperature or temperature
itself (i.e. in the presence of a slight temperature gradient and/or minute
compositional inhomogeneity across the system) the domain structure abruptly
sets in when the spontaneous polarization appears in the softest part of the
sample (i.e. the part with maximal $T_{c}$). This takes place when the
difference in $T_{c}$ in the parts of the sample is just $(10^{-3}-10^{-2})%
{\rm K}$ for the displacive systems, and even smaller, $\left(
10^{-5}-10^{-4}\right) ${\rm K,} for the order-disorder systems. The period
of the structure then grows linearly with lowering temperature and quickly
becomes {\em much larger} than the corresponding Kittel period.

This result does not depend on specific geometry assumed in the present
model. Indeed, if local $T_{c}=T_{c}(z)$ varies continuously, it can be
approximated by a piece-wise distribution of a sequence of ``slices''. Upon
cooling the system first looses stability in the softest part of thickness $%
l_{s},$ which is derived from the position of the boundary where local $%
T_{c}=0,$ with respect to a domain structure with fine period $\propto \sqrt{%
l_{s}}.$ The domains extend into the bulk of the system and become wider
with further cooling, since $l_{s}$ increases. In electroded sample there
will be no domain branching and domain walls would run straight across all
transformed slices. Otherwise, discontinuities would have resulted in very
strong depolarizing field. If the overall inhomogeneity is small, the
picture would obviously remain very similar to the two-slice model solved
above. The same arguments remain valid if the inhomogeneity were to have
more complex form/distribution in a sample. The novel feature of the present
effect in case of {\em ferroelectics} is that the depolarizing field appears
not due to surface charges, which are screened out by the electrodes, but
because of the bulk inhomogeneity. In the case of {\em ferroelastics,}
inhomogeneity in the sample results in the {\em bulk} {\em stresses} that
cause the splitting of the system into domains. In this case too the domain
wall would run straight through the soft part of the crystal, since the
discontinuities would result in large stray elastic stresses.

We have shown that a very tiny temperature gradient, or a slight
compositional inhomogeneity, etc., would result in practically any crystal
eventually splitting into domains no matter how high the quality of it is.
The rapid evolution of the domain pattern, found in the present paper, when
it starts from very fine domains at $T_{c}$, which then grow linearly with
temperature to very large sizes is similar to what have been reported in
Ref. \cite{nakatani85} for $\sim 1$mm thick TGS\ crystals. It would be very
interesting to perform controlled experiments for the domain structure close
to the second order phase transitions. One could check, in particular, the
basic assumption of the present theory that the electric fields (elastic
stresses) accompanying ferroelectric (ferroelastic) phase transitions even
in slightly inhomogeneous media are compensated by formation of the domain
structures rather than, for example, by screening of the electric field by
charge carriers in ferroelectrics. Further understanding of the domain
formation at phase transitions in real crystals is very important given
that many properties of ferroelectrics and ferroelastics, used in
applications, are mainly determined by the domain structures.

\end{document}